\documentclass{article}

\usepackage{color}
\usepackage{amsmath, amssymb, amsthm}

\usepackage[pdftex]{graphicx}
\usepackage{bm}

\newcommand{\RM}{\mathbb{R}}
\newcommand{\ZM}{\mathbb{Z}}

\newtheorem{theorem}{Theorem}
\newtheorem{definition}{Definition}
\newtheorem{lemma}{Lemma} 
\newtheorem{prop}{Proposition} 
\newtheorem{cor}{Corollary}

\newtheorem*{proof*}{Proof}

\newcommand{\xvec}{\ensuremath{\boldsymbol{x}}}

\newcommand{\evec}{\ensuremath{\boldsymbol{e}}}
\newcommand{\vvec}{\ensuremath{\boldsymbol{v}}} 
\newcommand{\uvec}{\ensuremath{\boldsymbol{u}}}
\newcommand{\kvec}{\ensuremath{\boldsymbol{k}}}
\newcommand{\wvec}{\ensuremath{\boldsymbol{w}}}

\begin{document}

\title{{\bf Ronkin/Zeta Correspondence}
\vspace{10mm}}

\author{Takashi KOMATSU \\
Department of Mathematics, \\
Hiroshima University, \\
Higashihiroshima, Hiroshima 789-8526, Japan \\
Math. Research Institute Calc for Industry \\
Minami, Hiroshima, 732-0816, Japan \\ 
e-mail: ta.komatsu@sunmath-calc.co.jp 
\\ \\
Norio KONNO \\
Department of Applied Mathematics, Faculty of Engineering \\ 
Yokohama National University \\
Hodogaya, Yokohama, 240-8501, Japan \\
e-mail: konno-norio-bt@ynu.ac.jp 
\\ \\
Iwao SATO \\ 
Oyama National College of Technology \\
Oyama, Tochigi, 323-0806, Japan \\ 
e-mail: isato@oyama-ct.ac.jp 
\\ \\
Kohei SATO$^{\ast}$ \\ 
Oyama National College of Technology \\
Oyama, Tochigi, 323-0806, Japan \\ 
e-mail: k-sato@oyama-ct.ac.jp
}

\date{\empty }

\maketitle

\vspace{45mm}
\noindent
\begin{small}
{\bf Corresponding author$^{\ast}$}: Kohei Sato, Oyama National College of Technology, Oyama, Tochigi, 323-0806, Japan, \\ 
e-mail: k-sato@oyama-ct.ac.jp

\end{small}


\vspace{20mm}

\par\noindent








%









\clearpage

\begin{abstract}
The Ronkin function was defined by Ronkin \cite{Ronkin} in the consideration of the zeros of almost periodic function. 
Recently, this function has been used in various research fields in mathematics, physics and so on. Especially in mathematics, it has a closed connections with tropical geometry, amoebas, Newton polytopes and dimer models (see \cite{GKZ}, for example). 

On the other hand, we have been investigated a new class of zeta functions for various kinds of walks including quantum walks 
by a series of our previous work on ``Zeta Correspondence". The quantum walk is a quantum counterpart of the random walk. 
In this paper, we present a new relation between the Ronkin function and our zeta function for random walks and quantum walks. 
Firstly we consider this relation in the case of one-dimensional random walks. 
Afterwards we deal with higher-dimensional random walks. 
For comparison with the case of the quantum walk, we also treat the case of one-dimensional quantum walks. 
Fortunately, the Laurent polynomials were obtained through those Ronkin functions. Finally, we describe some properties about them by using the terminology of tropical geometry.  
Our results bridge between the Ronkin function and the zeta function via quantum walks for the first time.

\end{abstract}

\vspace{10mm}

\begin{small}
\end{small}
\begin{small}
\par\noindent
{\bf Abbr. title:} Ronkin/Zeta Correspondence
\end{small}

\vspace{10mm}

\section{Introduction} 

We have been investigated a new class of zeta functions for many kinds of walks including the {\em quantum walk} (QW) and 
the {\em random walk} (RW) by a series of our previous work on ``Zeta Correspondence" in \cite{K1, K2, K3, K4, K5, K7, K8, K6}. 
The QW can be interpreted as a quantum counterpart of RW. As for QW, see \cite{Konno2008, ManouchehriWang, Portugal, Venegas} 
and as for RW, see \cite{Konno2009, Norris, Spitzer}, for examples. 

In Walk/Zeta Correspondence \cite{K2}, 
a walk-type zeta function was defined without use of the determinant expressions of zeta function 
of a graph $G$, and various properties of walk-type zeta functions of RW, correlated random walk (CRW) 
and QW on $G$ were studied. 
Also, their limit formulas by using integral expressions were presented. 

In \cite{K8}, Komatsu et al. introduced the logarithmic zeta function by using the above walk-type zeta function, and 
presented a relation between the Mahler measure and the logarithmic zeta function for QWs and RWs on on a finite torus. 
So we call this relationship ``Mahler/Zeta Correspondence" for short. 

The Mahler measure is closely related to the Ronkin function (see \cite{Ronkin}). 
Specially, in the dimer model, the the free energy with respect to its partition function is presented as the special value of the 
Ronkin function (see \cite{L, KOS, T}). 
The Ronkin function was defined by Ronkin \cite{Ronkin} in the consideration of the zeros of almost periodic function. 
It is known that the Ronkin function appears in areas of the ameba and the Newton convex hull (see \cite{GKZ}, for example).  

Our results bridge between the Ronkin function and the zeta function research fields via RWs and QWs for the first time.

The rest of this paper is organized as follows. 
In Section 2, we briefly review ``Walk/Zeta Correspondence" investigated in \cite{K2}. 
Furthermore, we deal with the logarithmic zeta function for QWs and RWs on a finite torus, and state explicit formulas for 
the logarithmic zeta functions of one-dimensional QWs and higher-dimensional RWs on a finite torus. 
In Section 3, we present the relation between the Ronkin function and RW on the finite torus $T^d_N $ and the relation between the Ronkin function and QW on the one-dimensional torus $T^1_N $. Moreover, we discuss some properties of the Laurent polynomials obtained from these Ronkin functions corresponding to RWs and QWs.

\section{Zeta functions due to RWs and QWs } 

For the convenience of readers, we give a brief overview of ``Walk/Zeta Correspondence" and the {\em logarithmic zeta function} in our previous work \cite{K2, K8}. These are the fundamental tools in this paper. 

\subsection{The walk-type zeta function} \label{WTZ}

First we introduce the following notation: $\mathbb{Z}$ is the set of integers, $\mathbb{Z}_{\ge}$ is the set of non-negative integers, 
$\mathbb{Z}_{>}$ is the set of positive integers, $\mathbb{R}$ is the set of real numbers, and $\mathbb{C}$ is the set of complex numbers. 
Moreover, $T^d_N$ denotes the {\em $d$-dimensional torus} with $N^d$ vertices, where $d, \ N \in \mathbb{Z}_{>}$. Remark that $T^d_N = (\mathbb{Z} \ \mbox{mod}\ N)^{d}$.

Following \cite{K2} in which Walk/Zeta Correspondence on $T^d_N$ was investigated, we treat our setting for $2d$-state discrete time walk with a nearest-neighbor jump on $T^d_N$.

The discrete time walk is defined by using a {\em shift operator} and a {\em coin matrix} which will be mentioned below. 
Let $f : T^d_N \longrightarrow \mathbb{C}^{2d}$. For $j = 1,2,\ldots,d$ and $\xvec \in T^d_N$, the shift operator $\tau_j$ 
is defined by $(\tau_j f)(\xvec) = f(\xvec-\evec_{j})$, 
where $\{ \evec_1,\evec_2,\ldots,\evec_d \}$ denotes the standard basis of $T^d_N$. This means that the walks are taken along the direction $\{ \pm\evec_1, \pm\evec_2, \ldots, \pm\evec_d \}$. Therefore, the {\it direction polytope} ${\rm DP}_d$ {\it of degree} $d$ can be defined as follows:
\[
{\rm DP}_d = {\rm Conv}(\pm\evec_1,\pm\evec_2,\ldots,\pm\evec_d),
\] 
where the symbol ${\rm Conv}(S)$ means the convex hull spanned by a subset $S\subset T^d_N$.
Moreover, we define the {\it direction graph} ${\rm DG}_d$ {\it of degree} $d$ as the dual graph of ${\rm DP}_d$. 

The {\em coin matrix} $A=[a_{ij}]_{i,j=1,2,\ldots,2d}$ 
is a $2d \times 2d$ matrix with $a_{ij} \in \mathbb{C}$ for $i,j =1,2,\ldots,2d$. 
If $a_{ij} \in [0,1]$ and $\sum_{i=1}^{2d} a_{ij} = 1$ for any $j=1,2, \ldots, 2d$, then the walk is a CRW. 
We should remark that, in particular, when $a_{i1} = a_{i2} = \cdots = a_{i 2d}$ for any $i=1,2, \ldots, 2d$, this CRW becomes a RW. 
If $A$ is unitary, then the walk is a QW. So our class of walks contains RW, CRW, and QW as special models. 

To describe the evolution of the walk, we decompose the $2d \times 2d$ coin matrix $A$ as
\begin{align*}
A=\sum_{j=1}^{2d} P_{j} A,
\end{align*}
where $P_j$ denotes the orthogonal projection onto the one-dimensional subspace $\mathbb{C}\eta_j$ in $\mathbb{C}^{2d}$. 
Here $\{\eta_1,\eta_2, \ldots, \eta_{2d}\}$ denotes a standard basis on $\mathbb{C}^{2d}$.

The discrete time walk associated with the coin matrix $A$ on $T^d_N$ is determined by the $2d N^d \times 2d N^d$ matrix
\begin{align}
M_A=\sum_{j=1}^d \Big( P_{2j-1} A \tau_{j}^{-1} + P_{2j} A \tau_{j} \Big).
\label{unitaryop1}
\end{align}
The state at time $n \in \mathbb{Z}_{\ge}$ and location $\xvec \in T^d_N$ can be expressed by a $2d$-dimensional vector:
\begin{align*}
\Psi_{n}(\xvec)= {}^T
\left[
\Psi^{1}_{n}(\xvec), \Psi^{2}_{n}(\xvec), \ldots , \Psi^{2d}_{n}(\xvec) 
\right]
\in \mathbb{C}^{2d},
\end{align*}
where $T$ is the transposed operator. For $\Psi_n : T^d_N \longrightarrow \mathbb{C}^{2d} \ (n \in \mathbb{Z}_{\geq})$,
Eq. (1) gives the evolution of the walk as follows.
\begin{align}
\Psi_{n+1}(\xvec) \equiv (M_A \Psi_{n})(\xvec)=\sum_{j=1}^{d}\Big(P_{2j-1}A\Psi_{n}(\xvec+\evec_j)+P_{2j}A\Psi_{n}(\xvec-\evec_j)\Big).
\end{align} 
This equation means that the walker moves at each step one unit to the $- x_j$-axis direction with matrix $P_{2j-1}A$ or one unit 
to the $x_j$-axis direction with matrix $P_{2j}A$ for $j=1,2, \ldots, d$. 

Moreover, for $n \in \ZM_{>}$ and $\xvec = (x_1, x_2, \ldots, x_d) \in T^d_N$, the $2d \times 2d$ matrix $\Phi_n (x_1, x_2, \ldots, x_d)$ is given by 
\begin{align*}
\Phi_n (x_1, x_2, \ldots, x_d) = \sum_{\ast} \Xi_n \left(l_1,l_2, \ldots , l_{2d-1}, l_{2d} \right),
\end{align*} 
where the $2d \times 2d$ matrix $\Xi_n \left(l_1,l_2, \ldots , l_{2d-1}, l_{2d} \right)$ is the sum of all possible paths in the trajectory 
of $l_{2j-1}$ steps $- x_j$-axis direction and  $l_{2j}$ steps $x_j$-axis direction and $\sum_{\ast}$ is the summation 
over $\left(l_1,l_2, \ldots , l_{2d-1}, l_{2d} \right) \in (\ZM_{\ge})^{2d}$ satisfying 
\begin{align*}
l_1 + l_2 + \cdots + l_{2d-1} + l_{2d} = n, \qquad x_j = - l_{2j-1} + l_{2j} \quad (j=1,2, \ldots, d).
\end{align*} 
Here we put
\begin{align*}
\Phi_0 (x_1, x_2, \ldots, x_d)
= \left\{ 
\begin{array}{ll}
I_{2d} & \mbox{if $(x_1, x_2, \ldots, x_d) = (0, 0, \ldots, 0)$, } \\
O_{2d} & \mbox{if $(x_1, x_2, \ldots, x_d) \not= (0, 0, \ldots, 0)$},
\end{array}
\right.
\end{align*}
where $I_n$ is the $n \times n$ identity matrix and $O_n$ is the $n \times n$ zero matrix. Then, for the walk starting from $(0,0, \ldots, 0)$, we obtain  
\begin{align*}
\Psi_n (x_1, x_2, \ldots, x_d) = \Phi_n (x_1, x_2, \ldots, x_d) \Psi_0 (0, 0, \ldots, 0) \qquad (n \in \mathbb{Z}_{\ge}).
\end{align*} 
We call $\Phi_n (\xvec) = \Phi_n (x_1, x_2, \ldots, x_d)$ {\em matrix weight} at time $n \in \mathbb{Z}_{\ge}$ and location $\xvec \in T^d_N$ starting 
from ${\bf 0} = (0,0, \ldots, 0)$. 
When we consider the walk on not $T^d_N$ but $\ZM^d$, we add the superscript ``$(\infty)$" to the notation like $\Psi^{(\infty)}$ and $\Xi^{(\infty)}$.

This type is {\em moving} shift model called {\em M-type} here. Another type is {\em flip-flop} shift model called {\em F-type} whose coin matrix is given by 
\begin{align*}
A^{(f)} = \left( I_{d} \otimes \sigma \right) A,
\end{align*} 
where $\otimes$ is the tensor product and
\begin{align*}
\sigma 
=
\begin{bmatrix}
0 & 1 \\ 
1 & 0 
\end{bmatrix} 
. 
\end{align*}
The F-type model is also important, since it has a central role in the Konno-Sato theorem \cite{KS}, for example. 
When we distinguish $A$ (M-type) from $A^{(f)}$ (F-type), we write $A$ by $A^{(m)}$.

The measure $\mu_n (\xvec)$ at time $n \in \mathbb{Z}_{\ge}$ and location $\xvec \in T^d_N$ is defined by
\begin{align*}
\mu_n (\xvec) = \| \Psi_n (\xvec) \|_{\mathbb{C}^{2d}}^p = \sum_{j=1}^{2d}|\Psi_n^{j}(\xvec)|^p,
\end{align*} 
where $\|\cdot\|_{\mathbb{C}^{2d}}^p$ denotes the standard $p$-norm on $\mathbb{C}^{2d}$. As for RW and QW, we take $p=1$ and $p=2$, respectively. Then RW and QW satisfy 
\begin{align*}
\sum_{\xvec \in T_N^d} \mu_n (\xvec) = \sum_{\xvec \in T_N^d} \mu_0 (\xvec), 
\end{align*} 
for any time $n \in \ZM_{>}$. 

To consider the zeta function, we use the Fourier analysis. 
To do so, we introduce the following notation: $\mathbb{K}_N = \{ 0,1, \ldots, N-1 \}$ and $\widetilde{\mathbb{K}}_N = \{ 0 ,2 \pi/N, \ldots, 2 \pi (N-1)/N \}$.

For $f : \mathbb{K}_N^d \longrightarrow \mathbb{C}^{2d}$, the Fourier transform of the function $f$, denoted by $\widehat{f}$, is defined by the sum
\begin{align}
\widehat{f}(\kvec) = \frac{1}{N^{d/2}} \sum_{\xvec \in \mathbb{K}_N^d} e^{- 2 \pi i \langle \xvec, \kvec \rangle /N} \ f(\xvec),
\end{align}
where $\kvec=(k_1,k_2,\ldots,k_{d}) \in \mathbb{K}_N^d$. Here $\langle \xvec,  \kvec \rangle$ is the canonical inner product 
of $\mathbb{R}^d$, i.e., $\langle \xvec,  \kvec \rangle = \sum_{j=1}^{d} x_j k_j$. 
Then we see that $\widehat{f} : \mathbb{K}_N^d \longrightarrow \mathbb{C}^{2d}$. 
Moreover, we should remark that 
\begin{align}
f(\xvec) = \frac{1}{N^{d/2}} \sum_{\kvec \in \mathbb{K}_N^d} e^{2 \pi i \langle \xvec, \kvec \rangle /N} \ \widehat{f}(\kvec),
\end{align}
where $\xvec =(x_1,x_2,\ldots,x_{d}) \in \mathbb{K}_N^d$. By using 
\begin{align}
\widetilde{k}_j = \frac{2 \pi k_j}{N} \in \widetilde{\mathbb{K}}_N, \quad \widetilde{\kvec}=(\widetilde{k}_1,\widetilde{k}_2,\ldots,\widetilde{k}_{d}) \in \widetilde{\mathbb{K}}_N^d, 
\end{align}
we can rewrite Eqs. (3) and (4) in the following way: 
\begin{align*}
\widehat{g}(\widetilde{\kvec}) 
= \frac{1}{N^{d/2}} \sum_{\xvec \in \mathbb{K}_N^d} e^{- i \langle \xvec, \widetilde{\kvec} \rangle} \ g(\xvec),
\qquad 
g(\xvec) 
= \frac{1}{N^{d/2}} \sum_{\widetilde{\kvec} \in \widetilde{\mathbb{K}}_N^d} e^{i \langle \xvec, \widetilde{\kvec} \rangle} \ \widehat{g}(\widetilde{\kvec}),
\end{align*}
for $g : \mathbb{K}_N^d \longrightarrow \mathbb{C}^{2d}$ and $\widehat{g} : \widetilde{\mathbb{K}}_N^d \longrightarrow \mathbb{C}^{2d}$. 
In order to take a limit $N \to \infty$, we introduced the notation given in Eq. (5). 
We should note that as for the summation, we sometimes write ``$\kvec \in \mathbb{K}_N^d$" instead of ``$\widetilde{\kvec} \in \widetilde{\mathbb{K}}_N^d$". 
From the Fourier transform and Eq. (5), we have
\begin{align*}
\widehat{\Psi}_{n+1}(\kvec)=\widehat{M}_A(\kvec)\widehat{\Psi}_n(\kvec),
\end{align*}
where $\Psi_n : T_N^d \longrightarrow \mathbb{C}^{2d}$ and $2d \times 2d$ matrix $\widehat{M}_A(\kvec)$ is determined by
\begin{align*}                          
\widehat{M}_A(\kvec)=\sum_{j=1}^{d} \Big( e^{2 \pi i k_j/N} P_{2j-1} A + e^{-2 \pi i k_j /N} P_{2j} A \Big). 
\end{align*}
By using notations in Eq. (5), we get 
\begin{align}                          
\widehat{M}_A(\widetilde{\kvec})=\sum_{j=1}^{d} \Big( e^{i \widetilde{k}_j} P_{2j-1} A + e^{-i \widetilde{k}_j} P_{2j} A \Big). 
\end{align}
Next we will consider the following eigenvalue problem for $2d N^d \times 2d N^d$ matrix $M_A$: 
\begin{align}
\lambda \Psi = M_A \Psi, 
\end{align}
where $\lambda \in \mathbb{C}$ is an eigenvalue and $\Psi (\in \mathbb{C}^{2d N^d})$ is the corresponding eigenvector. 
Noting that Eq. (7) is closely related to Eq. (2), we see that Eq. (7) is rewritten as 
\begin{align}
\lambda \Psi (\xvec) = (M_A \Psi)(\xvec) = \sum_{j=1}^{d}\Big(P_{2j-1} A \Psi (\xvec+\evec_j)+P_{2j} A \Psi (\xvec-\evec_j)\Big),
\end{align} 
for any $\xvec \in \mathbb{K}_N^d$. From the Fourier transform and Eq. (7), we obtain
\begin{align*}
\lambda \widehat{\Psi} (\kvec) = \widehat{M}_A (\kvec) \widehat{\Psi} (\kvec), 
\end{align*}
for any $\kvec \in \mathbb{K}_N^d$. 
Then the characteristic polynomials of $2d \times 2d$ matrix $\widehat{M}_A(\kvec)$ for fixed $\kvec (\in \mathbb{K}_N^d)$ is 
\begin{align}                          
\det \Big( \lambda I_{2d} - \widehat{M}_A (\kvec) \Big) = \prod_{j=1}^{2d} \Big( \lambda - \lambda_{j} (\kvec) \Big),
\end{align}
where $\lambda_{j} (\kvec)$ are eigenvalues of $\widehat{M}_A (\kvec)$. 
Similarly, the characteristic polynomials of $2d N^d \times 2d N^d$ matrix $\widehat{M}_A$ is 
\begin{align*}                          
\det \Big( \lambda I_{2d N^d} - \widehat{M}_A  \Big) = \prod_{j=1}^{2d} \prod_{\kvec \in \mathbb{K}_N^d} \Big( \lambda - \lambda_{j} (\kvec) \Big).
\end{align*}
Therefore, by taking $\lambda = 1/u$, we get the following key result.
\begin{align}                          
\det \Big( I_{2d N^d} - u M_A  \Big) 
= \det \Big( I_{2d N^d} - u \widehat{M}_A  \Big) = \prod_{j=1}^{2d} \prod_{\kvec \in \mathbb{K}_N^d} \Big( 1 - u \lambda_{j} (\kvec) \Big).
\end{align}
We should note that for fixed $\kvec (\in \mathbb{K}_N^d)$, eigenvalues of  $2d \times 2d$ matrix $\widehat{M}_A (\kvec)$ are expressed as 
\begin{align*}
{\rm Spec} ( \widehat{M}_A (\kvec)) = \left\{  \lambda_{j} (\kvec) \ | \ j = 1, 2, \ldots, 2d \right\}. 
\end{align*}
Moreover, eigenvalues of $2d N^d \times 2d N^d$ matrix not only $\widehat{M}_A$ but also $M_A$ are expressed as 
\begin{align*}
{\rm Spec} ( \widehat{M}_A) = {\rm Spec} (M_A)= \left\{  \lambda_{j} (\kvec) \ | \ j = 1, 2, \ldots, 2d, \ \kvec \in \mathbb{K}_N^d \right\}. 
\end{align*}

By using notations in Eq. (5) and Eq. (9), we see that for fixed $\kvec (\in \mathbb{K}_N^d)$,  
\begin{align}                          
\det \Big( I_{2d} - u \widehat{M}_A (\widetilde{\kvec}) \Big) = \prod_{j=1}^{2d} \Big( 1 - u \lambda_{j} (\widetilde{\kvec}) \Big).
\end{align}
Furthermore, Eq. (6) gives the following important formula.
\begin{align*}                          
\det \Big( I_{2d} - u \widehat{M}_A (\widetilde{\kvec}) \Big) = \det \left(I_{2d} - u \times \sum_{j=1}^{d} \Big( e^{i \widetilde{k}_j} P_{2j-1} A 
+ e^{-i \widetilde{k}_j} P_{2j} A \Big) \right). 
\end{align*}

In this setting, we define the {\em walk-type zeta function} by 
\begin{align}
\overline{\zeta} \left(A, T^d_N, u \right) = \det \Big( I_{2d N^d} - u M_A \Big)^{-1/N^d}.
\end{align}
We should remark that our walk is defined on the ``site" $\xvec (\in T^d_N)$. 
On the other hand, the walk in \cite{K1} is defined on the ``arc" (i.e., oriented edge). 
However, both of the walks are the same for the torus case. As for a more detailed information, see Remark 1 in \cite{HKSSjfa}, for instance.

By Eqs. (10), (11) and (12), we get 
\begin{align*}
\overline{\zeta} \left(A, T^d_N, u \right) ^{-1}
=
\exp \left[ \frac{1}{N^d} \sum_{\widetilde{\kvec} \in \widetilde{\mathbb{K}}_N^d} \log \left\{ \det \Big( I_{2d} - u \widehat{M}_A (\widetilde{\kvec}) \Big) \right\} \right].
\end{align*}
Sometimes we write $\sum_{\kvec \in \mathbb{K}_N^d}$ instead of $\sum_{\widetilde{\kvec} \in \widetilde{\mathbb{K}}_N^d}$. 
Noting $\widetilde{k_j} = 2 \pi k_j/N \ (j=1,2, \ldots, d)$ and taking a limit as $N \to \infty$, we show
\begin{align*}
\lim_{N \to \infty} \overline{\zeta} \left(A, T^d_N, u \right) ^{-1}
=
\exp \left[ \int_{[0,2 \pi)^d} \log \left\{ \det \Big( I_{2d} - u \widehat{M}_A \left( \Theta^{(d)} \right) \Big) \right\} d \Theta^{(d)}_{unif} \right],
\end{align*}
if the limit exists for a suitable range of $u \in \RM$. 
We should note that when we take a limit as $N \to \infty$, we assume that the limit exists throughout this paper. 
Here $\Theta^{(d)} = (\theta_1, \theta_2, \ldots, \theta_d) (\in [0, 2 \pi)^d)$ and $d \Theta^{(d)}_{unif}$ denotes the uniform measure on $[0, 2 \pi)^d$, that is,
\begin{align*}
d \Theta^{(d)}_{unif} = \frac{d \theta_1}{2 \pi } \cdots \frac{d \theta_d}{2 \pi }.
\end{align*}
Then the following result was obtained.

\begin{theorem}[Komatsu, Konno and Sato \cite{K2}]
\begin{align*}
\overline{\zeta} \left(A, T^d_N, u \right) ^{-1}
&= \exp \left[ \frac{1}{N^d} \sum_{\widetilde{\kvec} \in \widetilde{\mathbb{K}}_N^d} \log \left\{ \det \Big( F(\widetilde{\kvec}, u) \Big) \right\} \right],
\\
\lim_{N \to \infty} \overline{\zeta} \left(A, T^d_N, u \right) ^{-1}
&=
\exp \left[ \int_{[0,2 \pi)^d} \log \left\{ \det \Big( F \left( \Theta^{(d)}, u \right)  \Big) \right\} d \Theta^{(d)}_{unif} \right],
\end{align*}
where 
\begin{align*}
F \left( \wvec , u \right) = I_{2d} - u \widehat{M}_A (\wvec), 
\end{align*}
with $\wvec = (w_1, w_2, \ldots, w_d) \in \RM^d$.
\end{theorem}

\subsection{The logarithmic zeta function} 

We define the logarithmic zeta function on a finite torus, and state its properties. 
Let $T^d_N$ denotes the $d$-dimensional torus with $N^d$ vertices, and a $2d \times 2d$ matrix 
$A=[a_{ij}]_{i,j=1,2,\ldots,2d}$ the coin matrix of a dicrete time walk on $T^d_N $, where 
$a_{ij} \in \mathbb{C}$ for $i,j =1,2,\ldots,2d$. 

Now, we introduce the {\em logarithmic zeta function} as follows(see \cite{K8}). 
\begin{align}
{\cal L} \left( A, T_{\infty}^d, u \right) = \log \left[ \lim_{N \to \infty} \left\{ \overline{\zeta} \left(A, T_N^d, u \right)^{-1} \right\} \right].
\end{align}
Then, the second equation in Theorem 1 immediately gives
\begin{theorem}
\begin{align*}
{\cal L} \left( A, T_{\infty}^d, u \right)
=
\int_{[0,2 \pi)^d} \log \left\{ \det \Big( I_{2d} - u \widehat{M}_A \left( \Theta^{(d)} \right) \Big) \right\} d \Theta^{(d)}_{unif}.
\end{align*}
\end{theorem}
\noindent
Note that throughout this paper, we assume ``$u \in \RM$" for our logarithmic zeta function ${\cal L} \left( A, T_{\infty}^d, u \right)$. 
The range of $u \in \RM$ depends on the model determined by a coin matrix $A$.

Moreover, we define $C_r (A, T^d_N)$ by
\begin{align}
\overline{\zeta} \left(A, T^d_N, u \right) = \exp \left( \sum_{r=1}^{\infty} \frac{C_r (A, T^d_N)}{r} u^r \right).
\end{align}

Let ${\rm Tr} (A)$ denote the trace of a square matrix $A$. 
Then by definition of ${\rm Tr}$, the following result was shown in \cite{K2}.

\begin{theorem}[Komatsu, Konno and Sato \cite{K2}]
\begin{align*}
C_r (A, T^d_N) 
&
= \frac{1}{N^d} \sum_{\widetilde{\kvec} \in \widetilde{\mathbb{K}}_N^d} {\rm Tr} \left( \left( \widehat{M}_A (\widetilde{\kvec}) \right)^r \right),
\nonumber
\\
\lim_{N \to \infty} C_r (A, T^d_N) 
&
= \int_{[0,2 \pi)^d} {\rm Tr} \left( \left( \widehat{M}_A (\Theta^{(d)}) \right)^r \right) d \Theta^{(d)}_{unif}
= {\rm Tr} \left( \Phi_r ^{(\infty)} ({\bf 0}) \right).
\end{align*}
\end{theorem}

An interesting point is that $\Phi_r ^{(\infty)} ({\bf 0})$ is the return ``matrix weight" at time $r$ for the walk on not $T_N^d$ but $\ZM^d$. 
We should remark that in general ${\rm Tr} ( \Phi_r ^{(\infty)} ({\bf 0}) )$ is not the same as the return probability at time $r$ for QW and CRW, but for RW. 

Furthermore, we introduce
\begin{align*}
C_r (A, T^d_{\infty}) = \lim_{N \to \infty} C_r (A, T^d_N). 
\end{align*}
Therefore, by using the above equation, Theorem 2, and Eq. (14), we have
\begin{theorem}
\begin{align*}
{\cal L} \left( A, T_{\infty}^d, u \right)
= - \sum_{r=1}^{\infty} \frac{C_r (A, T^d_{\infty})}{r} \ u^r.
\end{align*}
\end{theorem}

From now on, we will present the result on only ${\cal L} \left( A, T_{\infty}^d, u \right)$ and $C_r (A, T^d_{\infty})$, 
since the corresponding expression for ``without $\lim_{N \to \infty}$" is the essentially same (see Theorems 1 and 3, for example).
  
Next, we consider a relation between the logarithmic zeta function ${\cal L} \left( A, T_{\infty}^d, u \right)$ 
for two-state QWs on the one-dimensional torus $T_N^1$.

First we deal with general walks including QWs on the one-dimensional torus $T^1_N$ 
whose $2 \times 2$ coin matrix $A^{(m)}$ (M-type) or $A^{(f)}$ (F-type) as follows: 
\begin{align*}
A^{(m)}  
=
\begin{bmatrix}
a_{11} & a_{12} \\ 
a_{21} & a_{22} 
\end{bmatrix} 
, \qquad 
A^{(f)} 
=  
\begin{bmatrix}
a_{21} & a_{22} \\ 
a_{11} & a_{12} 
\end{bmatrix}
,
\end{align*}
since
\begin{align*}
A^{(f)} = \left( I_1 \otimes \sigma \right) A^{(m)} = \sigma  A^{(m)} = 
\begin{bmatrix}
0 & 1 \\ 
1 & 0 
\end{bmatrix}
\begin{bmatrix}
a_{11} & a_{12} \\ 
a_{21} & a_{22} 
\end{bmatrix} 
.
\end{align*}
Set $k=k_1$ and $\widetilde{k} = \widetilde{k}_1$. In this case, we take 
\begin{align*}
P_{1} 
=
\begin{bmatrix}
1 & 0 \\ 
0 & 0 
\end{bmatrix}
, \qquad 
P_{2} 
=
\begin{bmatrix}
0 & 0 \\ 
0 & 1 
\end{bmatrix}
. 
\end{align*}
Thus we immediately get
\begin{align} 
\widehat{M}_{A^{(m)}} (\widetilde{k})
&= e^{i \widetilde{k}} P_{1} A^{(m)} + e^{-i \widetilde{k}} P_{2} A^{(m)} 
= 
\begin{bmatrix}
e^{i \widetilde{k}} a_{11} & e^{i \widetilde{k}} a_{12} \\ 
e^{-i \widetilde{k}}a_{21} & e^{-i \widetilde{k}} a_{22} 
\end{bmatrix} 
,
\\
\widehat{M}_{A^{(f)}} (\widetilde{k})
&= e^{i \widetilde{k}} P_{1} A^{(f)} + e^{-i \widetilde{k}} P_{2} A^{(f)} 
= 
\begin{bmatrix}
e^{i \widetilde{k}} a_{21} & e^{i \widetilde{k}} a_{22} \\ 
e^{-i \widetilde{k}}a_{11} & e^{-i \widetilde{k}} a_{12} 
\end{bmatrix} 
.
\end{align}
By these equations, we have
\begin{align*}
\det \Big( I_{2} - u \widehat{M}_{A^{(s)}} (\widetilde{k}) \Big) 
= 1 - {\rm Tr} \left(  \widehat{M}_{A^{(s)}} (\widetilde{k}) \right) u + \det \left(  \widehat{M}_{A^{(s)}} (\widetilde{k}) \right) u^2 \qquad (s \in \{m,f\}).
\end{align*}
Then the result given in \cite{K2} can be rewritten in terms of the logarithmic zeta function ${\cal L} \left( A, T_{\infty}^d, u \right)$ as follows: 

\begin{prop}
\begin{align*}
{\cal L} \left( A^{(s)}, T_{\infty}^1, u \right) 
= \int_0^{2 \pi} \log \left\{ 1 - {\rm Tr} \left(  \widehat{M}_{A^{(s)}}  (\theta) \right) u + \det \left(  \widehat{M}_{A^{(s)}} (\theta) \right) u^2 \right\} \frac{d \theta}{2 \pi},
\end{align*}
for $s \in \{m,f\}$. 
\label{kimarid1}
\end{prop} 

Remark that Proposition 1 is also obtained by Theorem 1.

From now on, we focus on QWs in one dimension. One of the typical classes of QWs for $2 \times 2$ coin matrix $A^{(m)}$ (M-type) or $A^{(f)}$ (F-type) is as follows: 
\begin{align*}
A^{(m)} 
=
\begin{bmatrix}
\cos \xi & \sin \xi  \\ 
\sin \xi & - \cos \xi  
\end{bmatrix} 
, \qquad 
A^{(f)} 
=  
\begin{bmatrix}
\sin \xi & - \cos \xi \\ 
\cos \xi & \sin \xi  
\end{bmatrix}
\qquad ( \xi \in [0, 2\pi)).
\end{align*}
When $\xi = \pi/4$, the QW becomes the so-called {\em Hadamard walk} which is one of the most well-investigated model in the study of QWs. 
Then the result given in \cite{K2} can also be rewritten in terms of the logarithmic zeta function like Proposition as follows: 

\begin{prop} 
\begin{align*}
{\cal L} \left( A^{(s)}, T_{\infty}^1, u \right) 
= \int_0^{2 \pi} \log \left( F^{(s)} \left( \theta, u \right) \right) \frac{d \theta}{2 \pi},
\end{align*}
for $s \in \{m,f\}$, where
\begin{align*}
F^{(m)} \left( w, u \right) 
&= 1 - 2 i \cos \xi \sin w \cdot u - u^2,
\\
F^{(f)} \left( w, u \right) 
&= 1 - 2 \sin \xi \cos w \cdot u + u^2.
\end{align*}

Moreover, 
\begin{align*}
C_{2l} (A^{(m)}, T^1_{\infty}) 
&= 2 l \left(-  \cos^2 \xi \right)^{l} \sum_{m=1}^l \frac{1}{m} {l-1 \choose m-1}^2 \left( - \tan^2 \xi \right)^{m}
\\
&= 2 l \left(-  \cos^2 \xi \right)^{l-1} (\sin^2 \xi) \ {}_2F_1 \left( 1-l , 1-l ; 2 ; - \tan^2 \xi \right),
\\
C_{2l} (A^{(f)}, T^1_{\infty}) 
&= 2 l \left(\sin \xi \right)^{2l} \sum_{m=1}^l \frac{1}{m} {l-1 \choose m-1}^2 \left( - \cot^2 \xi \right)^{m}
\\
&= 2 l \left(\sin \xi \right)^{2(l-1)} (- \cos^2 \xi) \ {}_2F_1 \left( 1-l , 1-l ; 2 ; - \cot^2 \xi \right),
\\
C_{2l-1} (A^{(s)}, T^1_{\infty}) 
&= 0 \qquad (s \in \{m,f\}), 
\end{align*}
for $l=1,2, \ldots$ and $\xi \in (0,\pi/2).$ 
\end{prop} 

Note that the result on ${\cal L} \left( A^{(s)}, T_{\infty}^1, u \right)$ in Proposition 2 is also derived from Proposition 1. 
Here the following result holds.

\begin{theorem}[Komatsu, Konno, Sato and Tamura \cite{K8}]
Let $c^{(m)} = \sec \xi \cdot \left( u - u^{-1} \right)$ and $c^{(f)} = - \ {\rm cosec} \ \xi \cdot \left( u + u^{-1} \right)$. Then we have
\[ 
{\cal L} \left( A^{(m)}, T_{\infty}^1, u \right) 
= \log \left( \frac{ 1 - u^{2} + \sqrt{1 + 2 \cos (2 \xi) u^2 + u^{4}}}{2} \right)
\]
for $\xi \in (0, \pi/2)$ and $u \in (\cos \xi - \sqrt{\cos^2 \xi +1}, 0)$.
\[
{\cal L} \left( A^{(f)}, T_{\infty}^1, u \right) 
= \log \left( \frac{ 1 + u^{2} + \sqrt{1 + 2 \cos (2 \xi) u^2 + u^{4}}}{2} \right)
\] 
for $\xi \in (0, \pi/2)$ and $u \in (- \infty, 0)$.
\end{theorem}


Now, we consider a relation between the logarithmic zeta function ${\cal L} \left( A, T_{\infty}^d, u \right)$ 
for RWs on the higher-dimensional torus $T_N^1$, see \cite{K8}. 

From definition of the (simple symmetric) RW on $T^d_N$ (see \cite{Norris, Spitzer}), we easily see that 
\begin{align} 
{\rm Spec} \left( P^{(D,c)} \right) 
&= \left\{ \frac{1}{d} \sum^d_{j=1} \cos \left( \frac{2 \pi k_j }{N} \right) \bigg| \ k_1 , \ldots , k_d \in \mathbb{K}_N \right\} 
\nonumber
\\
&= \left\{ \frac{1}{d} e^{(d, \cos)} (\widetilde{\kvec}) \bigg| \ k_1 , \ldots , k_d \in \mathbb{K}_N \right\},
\end{align}
where $P^{(D,c)}$ is the transition probability matrix of the (simple symmetric) RW on $T^d_N$. 
Here the RW on $T^d_N$ jumps to each of its nearest neighbors with equal probability $1/(2d)$. 
Noting Eq. (17), the result given in \cite{K8} can be rewritten in terms of the logarithmic zeta function as follows:

\begin{prop}
\begin{align*}
{\cal L} \left( A_{RW}, T_{\infty}^d, u \right)
= \int_{[0,2 \pi)^d} \log \Bigg\{ F_{RW} \left( \Theta^{(d)}, u \right) \Bigg\} d \Theta^{(d)}_{unif},
\end{align*}
where 
\begin{align*}
F_{RW} \left( \wvec, u \right) = 1 -\frac{e^{(d, \cos)}(\wvec)}{d} \cdot u.
\end{align*}
Moreover, we have
\begin{align*}
C_{r} (A_{RW},T_{\infty}^d)
&= \int_{[0,2 \pi)^d} G_{RW} \left( \Theta^{(d)} \right) d \Theta^{(d)}_{unif}, 
\end{align*}
where 
\begin{align*}
G_{RW} \left( \wvec \right) = \left( \frac{e^{(d, \cos)} (\wvec)}{d} \right)^r.
\end{align*} 
\end{prop}

In particular, when $d=1$ and $d=2$, we have

\begin{prop} 
\begin{align*}
{\cal L} \left( A_{RW}, T_{\infty}^1, u \right)
&= \log \left( \frac{1+\sqrt{1-u^2}}{2} \right)
= - \sum_{n=1}^{\infty} B_{2n} \frac{u^{2n}}{2n},
\\
{\cal L} \left( A_{RW}, T_{\infty}^2, u \right)
&= 
- \frac{u^2}{8} \ {}_4 F_3 \left( \frac{3}{2}, \frac{3}{2}, 1, 1 ; 2, 2, 2 ; u^2 \right) = - \sum_{n=1}^{\infty} \left( B_{2n} \right)^2 \frac{u^{2n}}{2n},
\end{align*}
for $u \in (-1,0)$, where 
\begin{align*}
B_{2n} = {2n \choose n} \left( \frac{1}{2} \right)^{2n}.
\end{align*}
\end{prop}

\section{Ronkin/Zeta Correspondence} 

The Ronkin function was defined by Ronkin in \cite{Ronkin}, and is defined for a Laurent polynomial. Let 
\[
P=P(x_1 , \ldots , x_k) = \sum_{i_1 , \ldots , i_k \in \mathbb{Z}} a_{i_1, \ldots , i_k} x^{i_1 }_1 \cdots x^{i_k }_k 
\]
be a Laurent polynomial with only a finite number of the $a_{i_1, \ldots , i_k}$'s being non-zero. 
Then the {\em Ronkin function} of $P$ is defined as follows: 
\[
R(x_1 , \ldots , x_k )= \oint_{|z_1 |=1} \cdots \oint_{|z_k |=1} \log |P( e^{x_1 } z_1 , \ldots , e^{x_k } z_k )| 
\frac{d z_1 }{2 \pi i z_1 } \cdots \frac{d z_k }{2 \pi i z_k }. 
\]

In this section, we discuss a correspondence between the logarithmic zeta function introduced in the previous section and
the Ronkin function.

\subsection{Correspondence for RW} 

For the logarithmic zeta function of the one-dimensional RW, the following result follows. 

\begin{prop} 
Let 
\[
P_1( e^x z)=1- \frac{1}{2} \left( e^x z+ e^x \frac{1}{z} \right)u
\] 
and 
\[
R(x)= \oint_{|z|=1} \log |P_1( e^x z)| \frac{dz}{2 \pi i z}.  
\]
Then 
\[
{\cal L} (A_{RW} , T^1_{\infty} , u)=R(0). 
\]
\end{prop}

{\bf Proof}.  Substituting $d=1$ in Proposition 3, we obtain 
\begin{align*}
{\cal L} (A_{RW} , T^1_{\infty} , u)= \int^{2 \pi}_0 \log (1- \cos \theta u) \frac{d \theta }{2 \pi }. 
\end{align*}

Now, let 
\[
P_1( e^x z)=1-\frac{1}{2} \left( e^x z+ e^x \frac{1}{z} \right)u. 
\]
If we set $z= e^{i \theta } $, then we have 
\[
P_1( e^x z)=1- \cos \theta \cdot e^x u. 
\] 
Thus, we obtain 
\[
R(x)= \oint_{|z|=1} \log |P( e^x z)| \frac{dz}{2 \pi i z} 
= \int^{2 \pi}_0 \log |1- \cos \theta \cdot e^x u| \frac{d \theta }{2 \pi }. 
\]

If $| e^x u|<1$, then $1- \cos \theta \cdot e^x u>0$, and so 
\[
R(x)= \int^{2 \pi}_0 \log (1- \cos \theta \cdot e^x u) \frac{d \theta }{2 \pi }. 
\] 
Thus, we have 
\[
R(0)= \int^{2 \pi}_0 \log (1- \cos \theta \cdot u) \frac{d \theta }{2 \pi }. 
\] 
Therefore, it follows that 
\[
{\cal L} (A_{RW} , T^1_{\infty} , u)=R(0). 
\]
$\Box$

Next, the relation between the Ronkin function and the 2-dimensional RW is given as follows.

\begin{prop} 
Let 
\begin{align*}
P_2( e^{x_1} z_1 , e^{x_2} z_2 )=1- \frac{1}{4} \left\{ \left( e^{x_1} z_1 + e^{x_1} \frac{1}{z_1} \right)
+\left( e^{x_2} z_2 + e^{x_2} \frac{1}{z_2} \right) \right\}  u
\end{align*}
and 
\[
R(x_1 , x_2 )= \oint_{|z_1 |=1} \oint_{|z_2 |=1} \log |P_2( e^{x_1 } z_1, e^{x_2 } z_2 )| 
\frac{dz_1}{2 \pi i z_1} \frac{dz_2}{2 \pi i z_2 }.  
\]
Then 
\[
{\cal L} (A_{RW} , T^2_{\infty} , u)=R(0,0). 
\]
\end{prop}

{\bf Proof}.  Substituting $d=2$ in Proposition 3, we obtain 
\[
{\cal L} (A_{RW} , T^2_{\infty} , u)= \int^{2 \pi}_0 \int^{2 \pi}_0 \log \left(1- \frac{1}{2} ( \cos \theta {}_1 + 
\cos \theta {}_2 )u\right) \frac{d \theta {}_1}{2 \pi } \frac{d \theta {}_2 }{2 \pi }. 
\]
If $-1<u<1$, then we have 
\[
{\cal L} (A_{RW} , T^2_{\infty} , u)= \int^{2 \pi}_0 \int^{2 \pi}_0 \log \left|1- \frac{1}{2} ( \cos \theta {}_1 + 
\cos \theta {}_2 )u\right| \frac{d \theta {}_1}{2 \pi } \frac{d \theta {}_2 }{2 \pi }. 
\]

Now, let 
\[
P_2( e^{x_1 } z_1 , e^{x_2 } z_2 )=1- \frac{1}{4} \left\{ \left( e^{x_1 } z_1 + e^{x_1 } \frac{1}{z_1} \right)
+\left( e^{x_2 } z_2 + e^{x_2 } \frac{1}{z_2} \right) \right\} u.
\] 
If we set $z_1 = e^{i \theta {}_1 } $ and $z_2 = e^{i \theta {}_2 } $, then we have 
\[
P_2( e^{x_1 } z_1 , e^{x_2 } z_2 )=1- \frac{1}{2} ( \cos \theta {}_1 + \cos \theta {}_2 )u.
\] 
Thus, we obtain 
\begin{align*} 
R(x_1 , x_2 )
&= \oint_{|z_1 |=1} \oint_{|z_2 |=1} \log |P_2( e^{x_1 } z_1 , e^{x_2 } z_2 )| 
\frac{d z_1 }{2 \pi i z_1 } \frac{d z_2 }{2 \pi i z_2 } 
\\ 
&= \int^{2 \pi}_0  \int^{2 \pi}_0 \log \left|1- \frac{1}{2} ( e^{x_1 } \cos \theta {}_1 +  e^{x_2 } \cos \theta {}_2 )u\right| 
\frac{d \theta {}_1 }{2 \pi } \frac{d \theta {}_2 }{2 \pi }. 
\end{align*}
Therefore, it follows that 
\[
R(0,0)= \int^{2 \pi}_0 \int^{2 \pi}_0 \log \left|1- \frac{1}{2} ( \cos \theta {}_1 + \cos \theta {}_2 )u\right| 
\frac{d \theta {}_1 }{2 \pi } \frac{d \theta {}_2 }{2 \pi }. 
\]
Hence, 
\[
{\cal L} (A_{RW} , T^2_{\infty} , u)=R(0,0). 
\]
$\Box$

Similarly to the proofs of Propositions 5 and 6, we obtain the following result.

\begin{theorem} 
Let $d \geq 2$. 
Furthermore, let 
\[
P_d=P_d( e^{x_1} z_1 , \ldots , e^{x_d} z_d)=1- \frac{u}{2d} \sum^d_{j=1} \left( e^{x_j } z_j + e^{x_j } \frac{1}{z_j} \right)  
\]
and 
\[
R(x_1 , \ldots , x_d )= \oint_{|z_1 |=1} \cdots \oint_{|z_d |=1} \log \left|1- \frac{1}{2d} \sum^d_{j=1} \left( e^{x_j } z_j + e^{x_j } \frac{1}{z_j} \right)\right| 
\frac{d z_1 }{2 \pi i z_1 } \cdots \frac{d z_d }{2 \pi i z_d }. 
\]
Then 
\[
{\cal L} (A_{RW} , T^d_{\infty } ,u)= R(0, \ldots ,0) \qquad (-1<u<1). 
\]
\end{theorem}

\subsection{Correspondence for QW} 

We consider a relation betweeen the logarithmic zeta function ${\cal L} (A, T^1_{\infty } ,u)$ and the Ronkin function 
for two-state QWs on the one-dimensional torus $T^1_n $. 
We deal with general walks including QWs on the one-dimensional torus $T^1_N$ 
whose $2 \times 2$ coin matrix $A^{(m)}$ (M-type) or $A^{(f)}$ (F-type) as follows: 
\begin{align*}
A^{(m)}  
=
\begin{bmatrix}
a_{11} & a_{12} \\ 
a_{21} & a_{22} 
\end{bmatrix} 
, \qquad 
A^{(f)} 
=  
\begin{bmatrix}
a_{21} & a_{22} \\ 
a_{11} & a_{12} 
\end{bmatrix}
. 
\end{align*}

We only state the case of the M-type $A=A^{(m)} $. 
For the case of the F-type, the result follows similarly.

\begin{theorem} 
Let 
\[
P^{(m)} =P^{(m)} ( e^x z) =1- \cos \xi \cdot \left(z- \frac{1}{z} \right) \cdot e^x u- u^2 
\]
and 
\[
P^{(f)} =P^{(f)} ( e^x z) =1- \sin \xi \cdot \left(z+ \frac{1}{z} \right) \cdot e^x u+ u^2. 
\]
Furthermore, let 
\[ 
R^{(s)} (x)= \oint_{|z|=1} \log |P^{(s)} ( e^x z)| \frac{dz}{2 \pi i z}. 
\]
Then 
\[
{\cal L} (A^{(s)} , T^d_{\infty } ,u)= R^{(s)} (0) \qquad (s \in \{ m,f \}). 
\]
\end{theorem}

{\bf Proof}.  At first, let $-1<u<1$. 
By Proposition 2, we have 
\[
{\cal L} ( A^{(m)} , T^1_{\infty } u)= \int^{2 \pi}_0 \log \left(1-2i \cos \xi \cdot \sin \theta \cdot u- u^2 \right) \frac{d \theta }{2 \pi }.
\]
By Theorem 5, we get 
\begin{align*}
{\cal L} ( A^{(m)} , T^1_{\infty } u)= \log \left( \frac{1- u^2 \sqrt{1+2 \cos (2 \xi ) u^2 + u^4}}{2} \right). 
\end{align*} 
Here, we use $-1<u<1$ in the the second equality.

Next, we obtain  
\begin{align*}  
R^{(m)} (0) &= \oint_{|z|=1} \log |P^{(s)} (z)| \frac{dz}{2 \pi i z} \\
&= \int^{2 \pi}_0 \log |1-2i \cos \xi \cdot \sin \theta \cdot u - u^2 | \frac{d \theta }{2 \pi }
\\ 
&= \frac{1}{2} \int^{2 \pi}_0 \log ( |1-2i \cos \xi \cdot \sin \theta \cdot u - u^2 |^2 ) \frac{d \theta }{2 \pi }
\\ 
&= \frac{1}{2} \int^{2 \pi}_0 \log ((1- u^2 )^2 +4 \cos {}^2 \xi \cdot \sin {}^2 \theta \cdot u^2 ) \frac{d \theta }{2 \pi }
\\ 
&= \frac{1}{2} \int^{2 \pi}_0 \log (1-2 \sin {}^2 \xi \cdot u^2 + u^4 -2 \cos {}^2 \xi \cdot \cos (2 \theta ) \cdot u^2 ) \frac{d \theta }{2 \pi }
\\ 
&= \frac{1}{2} \int^{2 \pi}_0 \log (1-2 \sin {}^2 \xi \cdot u^2 + u^4 -2 \cos {}^2 \xi \cdot \cos \theta \cdot u^2 ) \frac{d \theta }{2 \pi }
\\ 
&= \frac{1}{2} \int^{2 \pi}_0 \log (a-b \cos \theta ) \frac{d \theta }{2 \pi },  
\end{align*} 
where 
\[
a=1-2 \sin {}^2 \xi \cdot u^2 + u^4,\quad b= 2 \cos {}^2 \xi \cdot u^2. 
\]
If $-1<u<1$, then we have 
\[
a>0, \ b>0. 
\]
Furthermore, it holds that 
\[
\left|\frac{-b}{a} \right|= \frac{b}{a} \leq 1. 
\]
Thus, by (29) in \cite{K8}, we get 
\begin{align*}  
R^{(m)} (0) &= \frac{1}{2} \log a + \frac{1}{2} \log \left( \frac{1+ \sqrt{1- ( \frac{b}{a} )^2 } }{2} \right) 
\\ 
&= \frac{1}{2} \log \left( \frac{a+ \sqrt{ a^2 - b^2} }{2} \right). 
\end{align*}

Since 
\[
a^2 - b^2 = (1-u^2 )2 (1+ \cos (2 \xi ) u^2 + u^4),  
\]
we obtain the following result: 
\[
R(0)= {\cal L} ( A^{(m)} , T^1_{\infty } , u). 
\] 
$\Box$ 

\subsection{Properties of the Laurent polynomials due to RWs and QWs} 

In the previous sections, we introduced the Laurent polynomials:
\[
P_d( e^{x_1} z_1 , \ldots , e^{x_d} z_d)=1- \frac{u}{2d} \sum^d_{j=1} \left( e^{x_j } z_j + e^{x_j } \frac{1}{z_j} \right), 
\]
\[
P^{(m)} ( e^x z) =1- \cos \xi \cdot \left(z- \frac{1}{z} \right) \cdot e^x u- u^2,
\]
\[
P^{(f)} ( e^x z) =1- \sin \xi \cdot \left(z+ \frac{1}{z} \right) \cdot e^x u+ u^2. 
\]
From now on, we treat these functions as just Laurent polynomials, and discuss ``What these Laurent polynomials are?'' by using the terminologies of tropical geometry.

Since we do not need to consider restrictions on integral domains and so on, we assume $u=1$. 
 Furthermore, since we are only concerned with the zero set of these Laurent polynomials, we can replace $P_d$ and $P^{(s)}$ with the following:
\[
P_d( z_1 , \ldots , z_d)= \sum^d_{j=1} \left( z_j + \frac{1}{z_j} \right) -2d, 
\]
\[
P^{(m)} (z) = z- \frac{1}{z} - \frac{1}{\cos \xi},
\]
\[
P^{(f)} (z) = z+ \frac{1}{z} - \frac{1}{\sin \xi},
\]
where $\xi \neq \frac{n \pi}{2}, n\in {\mathbb Z}.$

The Newton polytope ${\rm NP}(P_d)\subset {\mathbb R}^d$ for the Laurent polynomial $P_d$ is the closed convex hull spanned by $\{ \pm\bm{e}_1, \pm\bm{e}_2, \ldots, \pm\bm{e}_d \}$, where $\{ \evec_1,\evec_2,\ldots,\evec_d \}$ is the standard basis of ${\mathbb R}^d$.
Especially, ${\rm NP}(P^{(m)}),\ {\rm NP}(P^{(f)})\subset {\mathbb R}$ are the closed section $[-1, 1]$. Fortunately, the Newton polytope ${\rm NP} (P_d)$ coincides with the direction polytope ${\rm DP}_d$ which is defined in the first part of subsection \ref{WTZ} as the following.
\begin{theorem}\label{direction1}
Let $P$ be one of the Laurent polynomials $P_d, P^{(m)}$ or $P^{(f)}$. Then, the Newton polytope ${\rm NP} (P)$ coincides with the direction polytope ${\rm DP}_d$ of its walk.
\end{theorem}
{\bf Proof}. 
If $P$ is $P^{(m)}$ or $P^{(f)}$, then the degree of the direction graph is one, and the following equation holds:
\[
{\rm DP}_d={\rm DP}_1={\rm Conv}(\evec_1, -\evec_1)=[-1, 1]={\rm NP}(P^{(m)})={\rm NP}(P^{(f)}).
\]
In the case that $P$ is $P_d$, the Newton polytope is ${\rm Conv}\{ \pm\bm{e}_1, \pm\bm{e}_2, \ldots, \pm\bm{e}_d \}$. Therefore, 
\[
{\rm DP}_d = {\rm Conv}\{ \pm\bm{e}_1, \pm\bm{e}_2, \ldots, \pm\bm{e}_d \} = {\rm NP} (P_d).
\]
$\Box$ 

Therefore, the Laurent polynomial $P$ contains the information concerning the direction of the walk. However, we conjecture that $P$ doesn't have any other infomation. In the following, we discuss the {\it amoebas} and {\it tropical hypersurfaces} of $P$ which contain more topological information compared with the Newton polytopes. 

For the Laurent polynomials $P^{(m)}$ and $P^{(f)}$, the discussion is omitted in the following because it is just a one-dimensional case and same with the case of $P_d$ for $d=1$.

The {\it amoeba} was first defined by Gelfand, Kapranov and Zelevinsky \cite{GKZ} as the image of the logarithmic map ${\rm Log}$ from the zero set $V(P)\subset (\mathbb{C}^{\times})^{d}$ of a Laurent polynomial $P$ to $\mathbb{R}^{d}$, 
$${\rm Log}:(z_1,\ldots, z_d)\to (\log|z_1|,\ldots,\log|z_d|).$$

By Passare and Rullg\aa rd \cite{PR}, the following assertions were shown.

\begin{prop}[Passare and Rullg\aa rd, \cite{PR}]\label{PR2}
The Ronkin function $R:\mathbb{R}^{d}\to\mathbb{R}$ is convex. Especially, it is strongly convex on the amoeba $\mathcal{A}$, and it is linear on each complement set $\mathbb{R}^{n}\setminus \mathcal{A}$.
\end{prop}

Forsberg, Passare and Tsikh \cite{FPT} have shown some relations between the structure of the amoeba ${\mathcal A}_P$ of $V(P)$ and the Newton polytope ${\rm NP}(P)$ by using the Ronkin function.

\begin{theorem}[Forsberg, Passare and Tsikh, \cite{FPT}]\label{FPT1}
The number of connected components of the amoeba complement $^{c}\mathcal{A}_P$ is at least equal to the number of vertices of the Newton polytope ${\rm NP}(P)$ and at most equal to the total number of integer points in ${\rm NP}(P)\cap {\mathbb Z}^d$.
\end{theorem}

\begin{figure}[hbtp]
 \begin{center}
 \includegraphics[scale=0.6]{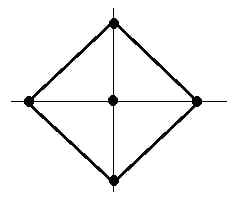}
 \caption{The Newton polytope ${\rm NP} (P_2)$}\label{Newton}
\end{center}
\end{figure}

For example, the amoeba ${\mathcal A}_{P_2}$ can be constructed as follows. 
\[
P_{2}( z_1 ,  z_2)= z_1 + \frac{1}{z_1} + z_2 + \frac{1}{z_2} -4.
\]
The Newton polytope ${\rm NP} (P_2)={\rm Conv}(\pm \evec_1, \pm \evec_2)$ is as Fig. \ref{Newton}. 

\begin{figure}[htbp]
    \centering
    \includegraphics[scale=0.7]{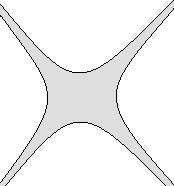}\hspace{50pt}  \includegraphics[scale=0.7]{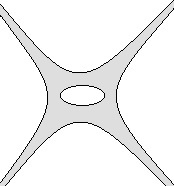}
    \caption{The amoeba $\mathcal{A}_{P_2}$}\label{Amoeba}
\end{figure}

By Proposition \ref{PR2} and Theorem \ref{FPT1}, it is shown that the number of the complement components in the amoeba $\mathcal{A}_{P_2}$ is equal to or less than the number of the lattice points in ${\rm Int}({\rm NP} (P_2))$, the amoeba $\mathcal{A}_{P_2}$ can be seen to be one of the figures in Fig. \ref{Amoeba}. In fact, the right figure is the amoeba of $P_2$ because the origin $(z_1, z_2)=(0, 0)$ is a pole of $P_2$. This area is sometimes called the ``bubble''.

By the results of Mikhalkin \cite{Mikhalkin} and Rullg\aa rd, it is known that the {\it ultra-discrete limit} of an amoeba converges in the Hausdorff metric to the {\it non-archimedean amoeba}. 
Moreover, it is proved that the non-archimedean amoeba coincides with the tropical hypersurface by Einsiedler, Kapranov and Lind \cite[Theorem 2.1.1]{EKL}. For a Laurent polynomial $P$, the tropical polynomial ${\rm Trop}(P)$ is given (see (\ref{tropicalpolynomial})), and the {\it tropical hypersurface} ${\mathcal T}_P$ of ${\rm Trop}(P)$ is defined as the set 
\[
\{ \vvec \in {\mathbb R}^d|\ \text{ the maximal in}\ {\rm Trop}(P)(\vvec) \text{ is achieved at least twice}\}.
\]

\begin{theorem}[Einsiedler, Kapranov and Lind, \cite{EKL}] \label{EKL1}
Let $P\neq 0$ be a Laurent polynomial. The non-archimedean amoeba $\mathcal{NA}_{P}$ coincides with the tropical hypersurface ${\mathcal T}_P$ for $P$.
\end{theorem}

In the following, we introduce non-archimedean amoeba, and give the tropical hypersurface for the Laurent polynomials $P_d$.

Let ${\mathbb K}=\bigcup_{n\geq1}{\mathbb C}((t^{1/n}))$ be the field of {\it Puiseux series} over ${\mathbb C}$, where ${\mathbb C}((t^{1/n}))$ is the field of Laurent polynomials in the formal variable $t^{1/n}$. For a Puiseux series:
\[
f=f(t)=\sum_{p\in I} a_p t^p,
\]
where $I\subset {\mathbb Q}$ and $a_p \neq 0$, the index set $I$ of $f$ is called the {\it support} of $f$, and denoted by $I_f$. This is a well-orderd set. A {\it valuation} on ${\mathbb K}$, $val: {\mathbb K} \rightarrow {\mathbb Q}\cup \{\infty \}$ can be defined as follows:
\[
val(f):={\rm min}\ I_f \ (f\neq0),\quad val(0)=\infty. 
\]

For instance, if $f\in {\mathbb C}$, then $f$can be represent as $f=f\cdot s^{0}$, and $I_f=\{ 0 \}$. Therefore, we have $val(f)=0$. 

\begin{definition}\upshape
A map $||*||:{\mathbb C}\rightarrow {\mathbb R}$ which satisfies the conditions

\begin{itemize}\centering
\item $||a||=0\ \Leftrightarrow \ a=0$, 
\item $||a\cdot b|| = ||a||\cdot ||b||$,
\item $||a+b|| \leq {\rm max}\{ ||a||,\ ||b|| \}$
\end{itemize}
for any $a,\ b\in {\mathbb C}$ is called a {\it non-archimedean norm} on ${\mathbb C}$.
\end{definition}

The map $||*||:{\mathbb C} \rightarrow {\mathbb R}$ given as following is a non-archimedean norm:
\[
||a||:= e^{-val(a)},\quad ||0||=0.
\] 
The image of the zero set $V(P_d)$ of $P_d$ by the map ${\rm Log}$ using this norm is called the {\it non-archimedean amoeba} of $P_d$.
\begin{equation}\label{Logmap}
{\rm Log}: ({\mathbb C^{\times}})^d \rightarrow {\mathbb R}^d,\quad (a_1, a_2,\ldots, a_d) \mapsto (\log||a_1||, \log||a_2||, \ldots, \log||a_d||).
\end{equation}
We note that $\log||a_i||=-val(a_i)$ for $1\leq i \leq d$.
\[
(\log||a_1||, \log||a_2||, \ldots, \log||a_d||) = (-val(a_1), -val(a_2), \ldots, -val(a_d)).
\]

By the logarithmic map in (\ref{Logmap}), a Laurent polynomial $P=\sum_{i, j} a_{i,j}(z_j)^i$ corresponds to the {\it tropical polynomial} of $P$ which is given as
\begin{equation}\label{tropicalpolynomial}
{\rm Trop}(P) = \bigoplus_{i,j} (i\cdot x_j -val(a_{i,j})),
\end{equation}
where $x_j = val(z_j)$ for $1\leq j \leq d$ is a variable over ${\mathbb R}$, and $\oplus$ means the max-plus (i.e., the additional operator $a\oplus b = {\rm max}\{ a, b\}$). This tropical polynomial is also called ``the {\it tropicalization} of $P$'' or ``the {\it ultra-discretization} of $P$''.

\begin{figure}[htbp]
    \centering
    \includegraphics[scale=1.3]{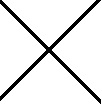}
    \caption{The tropical curve $\mathcal{T}_{P_2}$}\label{Tropical}
\end{figure}

As an example, the tropical curve given by $P_2$ is coincides with the non-archimedean amoeba of $P_2$ by Theorem \ref{EKL1}. The tropical polynomial corresponding to $P_2$ is as follows:
\[
{\rm Trop}(P_2) = x_1 \oplus (-x_1) \oplus x_2 \oplus (-x_2) \oplus 0 \equiv  x_1 \oplus (-x_1) \oplus x_2 \oplus (-x_2)
\]
where the symbol $\equiv$ means equality as functions. Then, the tropical curve given by ${\rm Trop}(P_{2})$ is as in Fig. \ref{Tropical}.

This tropical curve $\mathcal{T}_{P_2}$ is the ultra-discrete limit of $\mathcal{A}_{P_2}$. The bubble at the origin in the amoeba is deleted. This means that the topological information is lost by the ultra-discretization.


In the general case, for the Laurent polynomial
\[
P_d( z_1 , \ldots , z_k)= \sum^d_{j=1} \left( z_j + \frac{1}{z_j} \right) -2d, 
\]
and the Newton polytope ${\rm NP} (P_d)$ is
\[
{\rm NP} (P_d) ={\rm Conv}(\pm \evec_1, \ldots, \pm \evec_d).
\]
Furthermore, the tropical polynomial is
\[
{\rm Trop}(P_d)=\bigoplus_{i=1}^{d} (x_i \oplus (-x_i)) \oplus 0.
\]
Let $S\subset \{ x_i,\ -x_i|\ 1\leq i\leq d\}$ be a set of the coordinates and them multiplied by $-1$, where $S$ does not contain $x_i$ and $-x_i$ simultaneously. We set $\#S=r$ $(r\leq d)$ and
\[
S=\{ s_1,\ s_2,\ \ldots,\ s_r \}.
\]
The tropical hypersuface $\mathcal{T}_{P_d}$ consists of the $(d-r+1)$-dimensional strongly convex polyhedral cones:
\[
C_S := \sum_{x_j \notin S} {\mathbb R}_{\geq 0} \left( \left( \sum_{i=1}^{r}s_i \right) \pm x_j \right).
\]
This closed cone is the area such that the elements in $S$ are maximal in ${\rm Trop}(P_d)$, and corresponds to the $(r-1)$-dimansional face in ${\rm NP}(P_d)$:
\[
F_S:= \sum_{i=1}^{r}t_i s_i \text{ where} \sum_{i=1}^{r} t_i =1.
\]
Moreover, the defining equation of the $(d-r+1)$-dimensional hyperplane $H$ which contain the cone $C_S$ is given as
\[
s_1=s_2=\ldots =s_r,
\]
where the operator in this equation is the ordinary one (i.e., not tropical).

\begin{lemma}\label{perpen}
$C_S$ is perpendicular to $F_S$.
\end{lemma}
{\bf Proof}. 
In this proof, we identifies the coordinates $x_i$ with the standard basis $\evec_i$ for $1\leq i \leq d$.
The barycenter of the face $F_S$ is $\frac{1}{r}\sum_{i=1}^{r}s_i$. Therefore, a point in $F_S$ can be represent as
\[
\left( \sum_{i=1}^{r}t_i (s_1 - s_i ) +\dfrac{1}{r}\sum_{i=1}^{r}s_i \right)
\]
for integers $(t_1,\ t_2,\ \ldots,\ t_r)\in{\mathbb R}^{r}$. The direction vectors of $F_S$ are
\[
\uvec_i := s_1-s_i,\ (1\leq i \leq r).
\]
On the other hand, the vector in $C_S$ can be written as
\[
\vvec := \sum_{x_j \notin S} a_j \left( \left( \sum_{i=1}^{r}s_i \right) + x_j \right) + \sum_{x_j \notin S} \bar{a}_j \left( \left( \sum_{i=1}^{r}s_i \right) - x_j \right).
\] 
for non-negative integers $a_j, \bar{a_j}$. We note that the inner product $s_i\cdot s_j$ and $s_i\cdot x_j$ satisfy that
\[
s_i\cdot s_j=\left\{
\begin{array}{l}
1\ (i=j) \\
0\ (i\neq j)
\end{array}
\right.\quad \text{and}\quad  
s_i\cdot x_j=\left\{
\begin{array}{l}
\pm1\ (i=j) \\
0\ (i\neq j)
\end{array}
\right.  .
\] 
Therefore, we have
\[
\uvec_i\cdot \vvec=\sum_{x_j \notin S} a_j \left( \left( \sum_{i=1}^{r}s_i \right)\cdot (s_1 -s_i)\right) + \sum_{x_j \notin S} \bar{a}_j \left( \left( \sum_{i=1}^{r}s_i \right) \cdot (s_1 -s_i) \right)
\]
\[
=\sum_{x_j \notin S} a_j (s_1 \cdot s_1) - \sum_{x_j \notin S} a_j (s_i \cdot s_i) + \sum_{x_j \notin S} \bar{a_j} (s_1 \cdot s_1) - \sum_{x_j \notin S} \bar{a_j} (s_i \cdot s_i) = 0.
\]
$\Box$ 

By the definition of the cone $C_S$ and Lemma \ref{perpen}, the relation between ${\rm NP}(P_d)$ and ${\mathcal T}_{P_d}$ can be described as follows.

\begin{theorem}\label{duality1}
For the Laurent polynomial $P_d$, the tropical hypersurface ${\mathcal T}_{P_d}$ is given as the dual graph of the Newton polytope ${\rm NP}(P_d)$.
\end{theorem}

By Theorem \ref{duality1} and the definition of the ``direction graph'' in the first part of subsection \ref{WTZ}, we have the following conclusion.

\begin{cor}
Let $P$ be one of the laurent polynomials $P_d, P^{(m)}$ or $P^{(f)}$. Then, the tropical hypersurface ${\mathcal T}_{P_d}$ coincides with the direction graph ${\rm DG}_d$ of its walk.
\end{cor}
{\bf Proof}. 
Let $G^{*}$ be the dual graph of a graph $G$. By Theorem \ref{direction1} and Lemma \ref{duality1},
\[
{\mathcal T}_{P_d} ={\rm NP}^{*}(P_d) = {\rm DP}^{*}(P_d)={\rm DG}_d.
\]
$\Box$ 

In this case, the the tropical hypersurface ${\mathcal T}_{P_d}$ is determined by the Newton polytope ${\rm NP}(P_d)$ completely, and it is clear that both of them have the same information of the walk. However, the amoeba ${\mathcal A}_{P_d}$ is a little different from them because the amoeba contains a bubble at the origin. The Laurent polynomial $P_d$ has a pole at the origin. This bubble does not seem to make much sense for the walk, the characterization of this bubble by the language of RW or QW is a topic for future work. In summary, it turns out that the Laurent polynomial $P_d$ has almost only directional information of RW or QW.

\vspace{15pt}

{\bf Acknowledgments} 
 
The first author is supported by the JSPS Grant-in-aid for young scientisits No. 22K13959.

\end{document}